\documentclass[aps,prd,reprint,nofootinbib,longbibliography,showkeys]{revtex4-1}
\usepackage{blindtext}
\usepackage{mathtools}
\usepackage{xcolor}

\usepackage{mathrsfs}
\usepackage{braket}

\usepackage[utf8]{inputenc}
\usepackage[english]{babel}
\usepackage{hyperref}
\definecolor{mypink1}{rgb}{0.858, 0.188, 0.478}
\definecolor{mypink2}{RGB}{219, 48, 122}
\hypersetup{
    colorlinks=true,
    linkcolor=blue,
    filecolor=blue,      
    urlcolor=red,
    citecolor=blue,
}
\usepackage{color,colortbl}
\definecolor{LightCyan}{rgb}{0.88,1,1}
\usepackage{tensind}
\tensordelimiter{?}

\usepackage{graphicx}
\usepackage{hyperref}
\DeclareGraphicsExtensions{.bmp,.png,.jpg,.pdf}
\usepackage{amsmath}
\usepackage{courier}
\usepackage{verbatim}
\usepackage{amssymb}
\usepackage{amsfonts}
\usepackage{natbib}
\usepackage{soul}

\begin{document}
\title{Gravitational Wave Resonance in Ultra-Light Dark Matter Halos}
\author{Paola C. M. Delgado$^{1\dagger}$}

\affiliation{$^1$Faculty of Physics, Astronomy and Applied Computer Science,
Jagiellonian University, 30-348 Krakow, Poland}

\begin{abstract}
 
\textbf{Abstract}\\
Ultra-Light Dark Matter (ULDM) halos constituted by Ultra-Light Axions (ULAs) generate gravitational potentials that oscillate in time. In this paper I show these potentials interact with gravitational waves, resonantly amplifying them. For all ULA masses considered, the resonance in the solar region is currently negligible, while in a denser dark matter environment, which may arise in different scenarios, it might become significant. The frequency of the amplified gravitational wave is equal to the ULA mass in the case of first resonance band, which represents the most efficient scenario.\\

\textcopyright \hspace{1mm} Published by the American Physical Society under the terms of the Creative Commons Attribution 4.0 International license. Further distribution of this work must maintain attribution to the author(s) and the published article’s title, journal citation, and DOI. Funded by SCOAP3.

\end{abstract}

\maketitle

\section{Introduction}
The nature of dark matter represents one of the most intriguing open questions in Cosmology and Astrophysics. Although abundant evidence points to its existence, a detection was never made, leading to a wide range of possibilities with regard to its fundamental character \cite{ARBEY2021103865, Bergström_2009,10.21468/SciPostPhysLectNotes.48}. Cold Dark Matter (CDM) is currently used in the Standard Cosmological Model ($\Lambda{\rm CDM}$), which is very successful in describing the Universe on large scales. However, on sub-galactic scales there still remain incompatibilities between the CDM description and the observed data, as CDM predicts more structure on small scales than what we observe \cite{Primack:2009jr}. 
An interesting solution to this problem comes from Ultra-Light Dark Matter (ULDM), namely Ultra-Light Axions (ULAs) of mass $10^{-22}$eV as a dark matter candidate
\cite{Hu:2000ke,LEE2016166,Ferreira:2020fam}. Apart from that, different ULA masses can be obtained in the scope of string theory \cite{PhysRevD.81.123530} and might constitute the totality or a fraction of dark matter in the cases where $m\gtrsim 10^{-27}$ \cite{Hlozek:2014lca,FSiddharthaGuzmán_2000}. 

On the other hand, gravitational waves have been providing us with unprecedented tools to test our Universe and it is natural to ask how they can shed light into the dark matter problem. A number of different dark matter candidates have already been investigated by using gravitational wave physics \cite{Fornal:2023hri, Mosbech:2022nkk, Samanta:2021mdm, Ghoshal:2023fhh}, including ULDM \cite{Khmelnitsky:2013lxt, Aoki:2016kwl, Fujita:2020iyx, Aghaie:2023lan}.

In this work I focus on a peculiar property of ULDM halos constituted by ULAs - the time-oscillation of the generated gravitational potentials - to show its relation with gravitational wave resonance in the halo. The mechanism corresponds to narrow band parametric resonance, enhancing gravitational waves of frequencies equal to the ULA masses. 

The paper is organised as follows: section \ref{sechalo} presents the description of the ULDM halo, resulting in the expressions for the gravitational potentials. Section \ref{secfloquet} introduces the main ideas regarding the mechanism of parametric resonance and its Floquet analysis. Section \ref{sectiongwresonance} applies parametric resonance to the context of a gravitational wave in a ULDM halo. Finally, a summary of the results is presented in section \ref{secconc}. Natural units with $M_p=1$ are used throughout the text, except when explicitly said otherwise.

\section{Description of the ULDM Halo}\label{sechalo}
In this section I present the mathematical description of the ULDM halo, which leads to the expressions for the oscillating gravitational potentials. As shown in section \ref{sectiongwresonance}, this feature is of major importance for the occurrence of resonance.

We start from an almost Minkowski space-time given by
\begin{equation}\label{eqds}
    ds^2=-(1+2U)dt^2+(1-2\bar{U})(dx^2+dy^2+dz^2),
\end{equation}
as the expansion of the Universe is negligible on the scales considered. The quantities $U$ and $\bar{U}$ are the gravitational potentials generated by the ULDM halo, which are treated perturbatively. This assumption holds even inside the halo, where the condition $U, \bar{U}\ll 1$ is still satisfied.

The ULDM is represented by an ULA field 
\begin{equation}\label{axioneq}
    \phi(t)=\phi_0\cos{(m t)},
\end{equation}
where the spatial dependence of the amplitude $\phi_0$ was neglected at leading order \cite {Khmelnitsky:2013lxt, Aoki:2017ehb}. The oscillating frequency corresponds to the particle's energy, which can be approximated by the particle's mass $m$ in the non-relativistic limit. The corresponding energy-momentum tensor is 
\begin{equation}
    T_{\mu\nu}={\rm diag}(\rho,p,p,p),
\end{equation}
where
\begin{eqnarray}
    \rho&=&\frac{1}{2} m^2 \phi_0^2,\\
    p &=& -\rho \cos{(2mt)}.
\end{eqnarray}
The oscillating pressure leads to an oscillating contribution to the gravitational potentials $U$ and $\bar{U}$, in such a way that we can write
\begin{eqnarray}
    T &=& T_0+\delta T,\\
    U &=& U_0 + \delta U,\\
    \bar{U} &=& \bar{U}_0 + \delta\bar{U},\\
    R &=& R_0+\delta R,
\end{eqnarray}
where $T$ is the trace of the energy-momentum tensor, $X_0$ terms are time-independent and $\delta X$ terms oscillate in time. Both $U$ and $\bar{U}$ depend on space and time, where the spatial dependence is restricted to $U_0$ and $\bar{U}_0$. $R$ is the Ricci scalar associated to \eqref{eqds}, which reads
\begin{equation}\label{Riccieq}
R=-6\ddot{\bar{U}}+2\nabla^2(2\bar{U}-U),
\end{equation}
where dot denotes derivative with respect to $t$. 

Now let us explore Einstein's equations
\begin{equation}
    R_{\mu\nu}-\frac{1}{2} R g_{\mu\nu}=T_{\mu\nu},
\end{equation}
where $R_{\mu\nu}$ is the Ricci tensor, in order to relate some of these quantities.
First, from the traceless part of the $ij$ component we obtain 
\begin{equation}
    \bar{U}_0=U_0.
\end{equation}
Secondly, from the trace of the Einstein's equations, we have $-R=T$, which allows us to identify
\begin{equation}
    R_0 = \rho.
\end{equation}
On the other hand, from \eqref{Riccieq} we have the time-independent part of the Ricci scalar $R_0$ as a function of the gravitational potentials
\begin{equation}
    R_0=2\nabla^2 U_0,
\end{equation}
leading to the following Poisson equation
\begin{equation}\label{eqpoisson}
    2 \nabla^2U_0=\rho.
\end{equation}
We can estimate the magnitude of $U_0$ by switching to Fourier space
\begin{equation}
    U_0\propto \frac{\rho}{k_a^2},
\end{equation}
where $k_a$ is the wavenumber related to the ULA. Its value can be estimated from $k_a^2/m^2=v^2$, where $v\simeq 10^{-3}$ is a typical velocity in our Galaxy. 

The expression for the oscillating part $\delta\bar{U}$ is obtained from \eqref{Riccieq} by assuming $\ddot{\delta\bar{U}}\gg \nabla^2\delta\bar{U}$ and $\ddot{\delta\bar{U}}\gg \nabla^2\delta U$\footnote{This assumption is consistent with \eqref{axioneq}, where the spatial dependence of $\phi_0$ was neglected \cite{Khmelnitsky:2013lxt}.}, which allows us to write \cite{Aoki:2017ehb}
\begin{equation}
    \delta T = 6 \ddot{\delta\bar{U}}.
\end{equation}
By solving this equation we obtain
\begin{equation}\label{dU}
    \delta\bar{U}=\frac{\rho}{8m^2}\cos{(2 m t)}.
\end{equation}

In order to find $\delta U$ we follow \cite{Aoki:2016kwl} and perform a change of frames, which retains the dependence on spatial derivatives of $\delta U$ in the $ij$ component of the Einstein's equations. Then we switch back to the halo frame by setting the relative velocity between the frames to zero, which leads to
\begin{equation}\label{eqdU}
    \delta U = - \delta \bar{U}.
\end{equation}

\section{Parametric resonance}\label{secfloquet}
Let us now introduce the parametric resonance mechanism in order to present the main ideas explored in the gravitational wave amplification in section \ref{sectiongwresonance}. 

Parametric resonance is a classical phenomenon that an oscillator experiences when it receives a periodically varying contribution to its mass. The result is an exponential amplification of the oscillator's amplitude if its frequency is within the so called resonance bands, which depend on the frequency by which the mass varies.  

As an example, let us consider a generic oscillator $x(t)$ governed by the following equation of motion
\begin{equation}\label{eqoscill}
    \ddot{x}+Ax-2q\cos{(2t)}x=0,
\end{equation}
where $A$ and $q$ are parameters determined by details of the dynamics of $x(t)$ and dot denotes derivative with respect to $t$. This type of equation is known as Mathieu equation and can be investigated by means of the Floquet instability theory, which can be used to estimate the oscillator's amplification. The exponential growth is quantified by means of the so called Floquet exponent $\mu$, which appears in the solution as
\begin{equation}\label{floquetest}
    x(t)\propto \exp{(\mu t)}.
\end{equation}
It can be computed by using the fundamental matrix, i.e. a matrix representing the independent solutions \cite{Amin:2014eta}. The eigenvalues $\sigma^{\pm}$ of this matrix at $t_0+T$, where $t_0$ is the initial time and $T$ is the period of the oscillating function in \eqref{eqoscill}, are related to the Floquet exponent $\mu$ via 
\begin{equation}
    \mathbb{R}[\mu^{\pm}]=\frac{1}{T}\ln{|\sigma^{\pm}|},
\end{equation}
where the superscript $\pm$ represents the growing and decaying solutions to \eqref{eqoscill}. When $\mathbb{R}[\mu^{\pm}]>0$, $x(t)$ is exponentially amplified, as seen from \eqref{floquetest}. By following this procedure, we find the values of $A$ that correspond to the resonance bands and the dependence of $\mu$ on $q$, which is given by 
\begin{eqnarray}
    \mu\propto \left\{\begin{matrix} q~ &{\rm if}&~ A\subset (1-q,1+q)\\
 q^2~ &{\rm if}&~ A\subset (4-q^2,4+q^2)
\end{matrix}\right..
\end{eqnarray}

If $q$, i.e. the periodically varying contribution to the mass term, is small compared to $A$, the parametric resonance is said to happen in a narrow band. As we will see in section \ref{sectiongwresonance}, this is exactly the case of the gravitational wave resonance in the ULDM halo and, therefore, the first resonance band, i.e. $A\subset (1-q,1+q)$, is the most efficient when it comes to the amplification factor.

\section{Gravitational Wave Resonance in the ULDM Halo}\label{sectiongwresonance}
\begin{figure}
\begin{center}
\includegraphics[width=8.5cm]{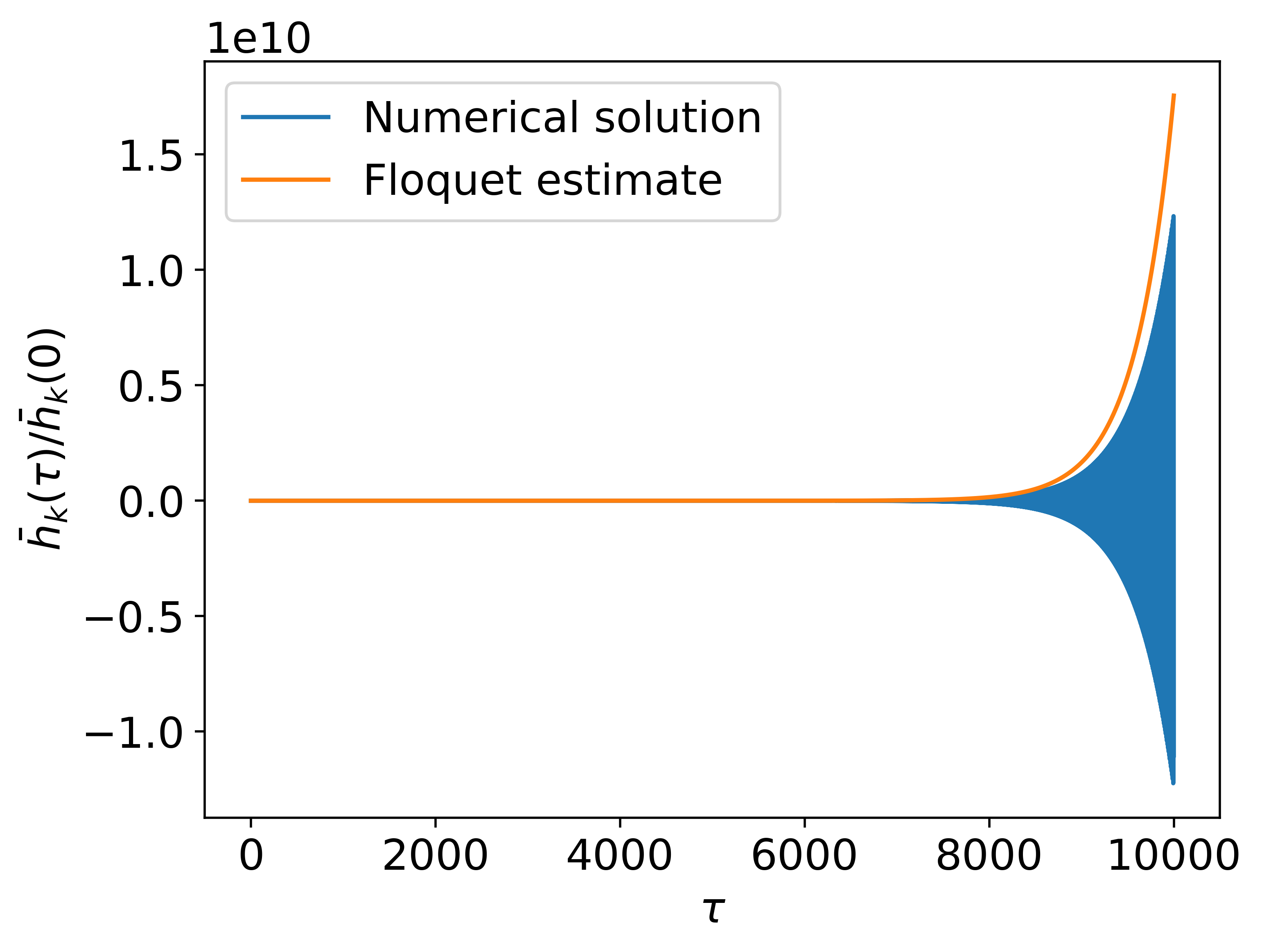}
\end{center}
\caption{\label{figamp} Gravitational wave resonance for $m=10^{-22}{\rm eV}$ and $\rho=10^{16} \times 0.4{\rm GeV/cm}^3$, where the factor $10^{16}$ was introduced to speed up the resonance and reduce computational time. For the real value in the solar region, $\rho=0.4{\rm GeV/cm}^3$, assuming the ULA constitutes $100\%$ of dark matter, the amplification becomes of $\mathcal{O}(1)$ around $\tau\sim 1/q \simeq 2.0\times 10^{18}$, equivalent to $3.9\times 10^{17}$ years. The Floquet estimate corresponds to \eqref{eqfloquet}. The initial conditions used are $h_k(0)=1$ and ${h_k}'(0)=0$.}
\end{figure}

Let us now consider a gravitational wave around the space-time \eqref{eqds}, which can be written as 
\begin{equation}
    h_{\mu\nu}=h \epsilon_{\mu\nu},
\end{equation}
where $h$ is the amplitude and $\epsilon_{\mu\nu}$ is the polarization tensor. Since the latter is parallel transported along the geodesics, we can neglect its change due to the presence of the gravitational potentials $U$ and $\bar{U}$ and write the equation of motion for the scalar amplitude $h$ \cite{Peters:1974gj, Grespan:2023cpa}
\begin{equation}
    \partial_\mu (\sqrt{-g} g^{\mu\nu}\partial_\nu h)=0,
\end{equation}
which leads to
\begin{eqnarray}\label{eqeom1}
 \nonumber   &&\ddot{h}-(1+2U+2\bar{U})\nabla^2 h-\dot{U}\dot{h}-3\dot{\bar{U}}\dot{h}+\\
    &&+\partial_i h \partial_i \bar{U}-\partial_i h\partial_i U=0
\end{eqnarray}
up to linear order in $U$ and $\bar{U}$.
By substituting \eqref{dU} and \eqref{eqdU} in \eqref{eqeom1}, we find, in Fourier space, 
\begin{eqnarray}
\nonumber   && \ddot{\bar{h}}_k+k^2 \bar{h}_k-4\int d^3\vec{x} \exp{(-i\vec{k}\cdot\vec{x})}U_0 \nabla^2\bar{h}+\\
&-&\frac{1}{2}\rho\cos{(2mt)}\bar{h}_k=0,
\end{eqnarray}
where $\bar{h}\equiv \exp{(\delta U)}h$. This field redefinition was performed in order to kill the friction terms present in \eqref{eqeom1}. Defining $\tau\equiv m t$, we obtain
\begin{eqnarray}
\nonumber   && {\bar{h}}''_k+\frac{k^2}{m^2} \bar{h}_k-\frac{4}{ m^2}\int d^3x \exp{(-i\vec{k}\cdot\vec{x})}U_0 \nabla^2\bar{h}+\\
&-&\frac{1}{2}\frac{\rho}{m^2}\cos{(2\tau)}\bar{h}_k=0,
\end{eqnarray}
where prime denotes derivative with respect to $\tau$. The remaining convolution is not trivial to perform, but it can be neglected if compared to the other kinetic term in the equation of motion. Because it is proportional to $(\rho/k_a^2) (k^2/m^2) \bar{h}_k$, it is very small compared to $(k^2/m^2) \bar{h}_k$, as  $\rho/k_a^2 \ll 1$. For this reason, we approximate 
\begin{eqnarray}
 \nonumber   A \bar{h}_k&\equiv& \frac{k^2}{m^2}\bar{h}_k-\frac{4}{m^2}\int d^3x \exp{(-i\vec{k}\cdot \vec{x})}U_0\nabla^2\bar{h}\\
    &\simeq& \frac{k^2}{m^2}\bar{h}_k,
\end{eqnarray}
which leads to
\begin{equation}\label{eqmathieu}
    {\bar{h}}''_k+A \bar{h}_k-2q\cos{(2\tau)} \bar{h}_k=0,
\end{equation}
where $q\equiv \rho/m^2/4$. Although the remaining convolution is not negligible compared to the oscillating term, they play different roles in the Mathieu equation. The first, together with the other kinetic term, establishes where the resonance bands are centered, while the second is related to the amplification factor and to the band width.
Therefore, equation \eqref{eqmathieu} is a Mathieu equation, just like \eqref{eqoscill}, in the first resonance band for $k^2=m^2$. Solving it numerically we obtain figure \ref{figamp}, which shows the amplification of the gravitational wave $\bar{h}_k$ in time $\tau$ for an illustrative high energy density. 

According to the Floquet instability theory, the solution to \eqref{eqmathieu} can be approximated as\footnote{Due to the fact that $\exp{(\delta U)}\simeq 1$ we can approximate $h_k\simeq \bar{h}_k$.}
\begin{equation}\label{eqfloquet}
    h_k\simeq \bar{h}_k\propto \exp{(q\tau/2)},
\end{equation}
which allows us to estimate the time required for the amplification to become of $\mathcal{O}(1)$, i.e. $\tau\sim 1/q$. It is important to note that $q$, which contains the information about the ULA energy density $\rho$, depends on the fraction of ULAs as dark matter $f$, as
\begin{equation}\label{fandrho}
    \rho=f \rho_{DM},
\end{equation}
where $\rho_{DM}$ is the dark matter energy density. 

\subsection{Gravitational wave resonance independently of constraints on ULDM}\label{noconst}
Because the gravitational wave resonance could, in principle, be used to independently test ULAs, I first assume $f=1$ for all masses, ignoring the constraints already imposed by other phenomena. In Section \ref{secwithconst} I present the results considering these constraints.

Let us first consider the solar region, i.e. $\rho=0.4 {\rm GeV/cm}^3$. In the most standard scenario, i.e. $m\simeq 10^{-22}{\rm eV}$, one would wait for $3.9\times10^{17}$ years to see an  amplification of gravitational waves with frequency in the Pulsar Timing Array (PTA) range, which is larger than the age of the Universe. For
ULAs of mass $m\simeq 10^{-27}{\rm eV}$, the required time is $3.9\times10^{12}$ years, still unfeasible for current tests. 

On the other hand, the prospects are largely improved if one considers very dense regions in the halos, although still satisfying $\rho/m^2\ll 1$, which can arise, for instance,  due to the existence of a black hole in the halo \cite{Chan:2022gqd, Nampalliwar:2021tyz, Banerjee:2019epw,Budker:2023sex}. For $\rho\simeq 1.4 \times 10^7$GeV/cm$^3$, the ULAs of mass $10^{-22}$eV would take $1.1\times10^{10}$ years to amplify gravitational waves, which is compatible with the time of formation of dark matter halos\footnote{This corresponds to the most conservative assumption and a bigger gravitational wave amplification is obtained if one considers scenarios where dark matter halos formed earlier in the Universe history \cite{Savastano:2019zpr, Hogan:1988mp, Fairbairn:2017sil}.}. In this case, primordial gravitational waves in the PTA range could be amplified if they keep traveling through this region since the halo formation. Shorter, and therefore more realistic, time intervals are achieved in denser environments. Note that for this specific energy density, $\rho/m^2\ll 1$ holds for $m\gtrsim  10^{-26}$.  

\begin{figure}
\begin{center}
\includegraphics[width=8.5cm]{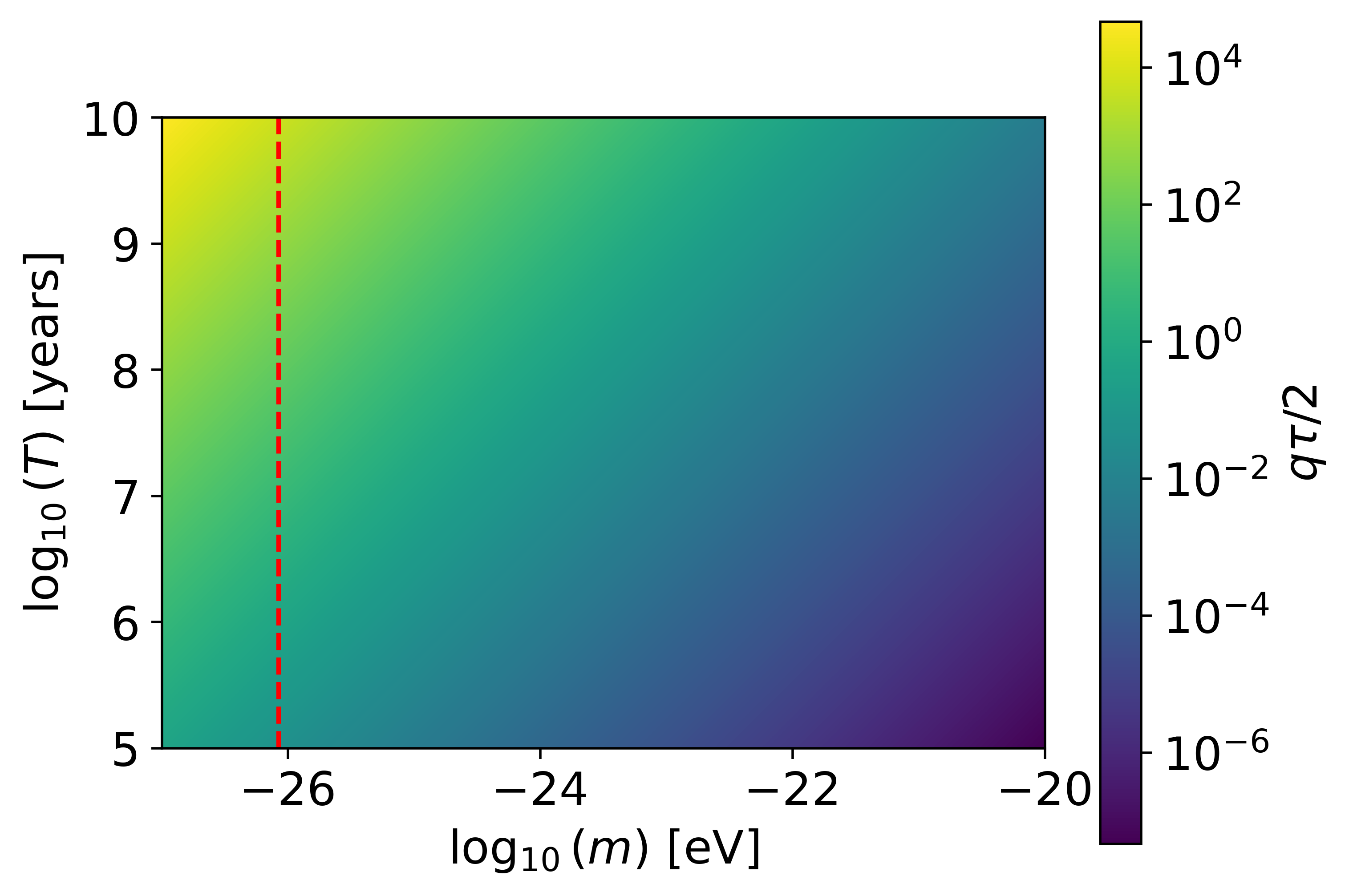}
\end{center}
\caption{\label{figampmasses} Argument of the amplification factor \eqref{eqfloquet}, $q\tau/2$, as a function of time and ULA masses assuming $f=1$ for all masses in the very dense dark matter region, i.e. $1.4\times 10^7$GeV/cm$^3$. The red line indicates when $\rho/m^2\simeq 0.01$, representing a left bound to the region where $\rho/m^2\ll 1$.}
\end{figure}

Figure \ref{figampmasses} depicts the amplification, represented by the argument of the exponential  function in \eqref{eqfloquet}, as a function of time and ULA masses. It is important to note that the formalism used in this work would break down when the condition $h\ll 1$ is not anymore satisfied, which sets an upper bound on the amplifications depending on the initial gravitational wave amplitude. The most interesting gravitational wave frequencies lie in the PTA range and correspond to $10^{-23}$eV$\lesssim m\lesssim 10^{-22}$eV, resulting in large time intervals for $\mathcal{O}(1)$ amplification, although still compatible with halo formation time. On the other hand, shorter time intervals are achieved by lighter ULAs, namely $10^{-26}{\rm eV}\lesssim m \lesssim 10^{-24}{\rm eV}$, which correspond to gravitational wave frequencies from $2.4\times 10^{-10}$Hz to $2.4\times 10^{-12}$Hz. Although not covered by ongoing gravitational wave detectors, these frequencies could be explored in the future \cite{Hisano:2019iqc, DeRocco:2022irl}. 

\subsection{Gravitational wave resonance considering constraints on ULDM}\label{secwithconst}
\begin{figure}
\begin{center}
\includegraphics[width=8.5cm]{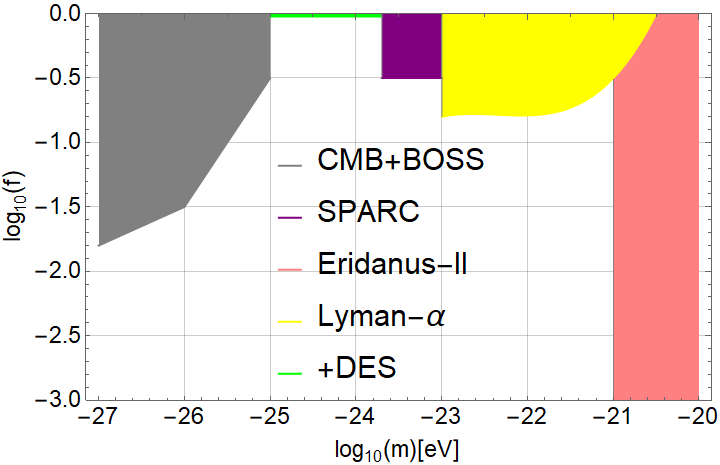}
\end{center}
\caption{\label{figconst}Constraints imposed to the ULA fraction as dark matter $f$ depending on the ULA mass $m$ for $10^{-27}{\rm eV}\gtrsim m \gtrsim 10^{-20}{\rm eV}$ \cite{Flitter:2022pzf}, namely CMB+BOSS \cite{Hlozek:2014lca,Lague:2021frh}, SPARC \cite{Bar:2021kti}, Eridanus-II \cite{Marsh:2018zyw}, Lyman-$\alpha$ forest \cite{Kobayashi:2017jcf} and galaxy weak lensing combined with Planck ($+$DES) \cite{Dentler:2021zij}. }
\end{figure}

Now let us consider constraints already imposed to the ULA fraction as dark matter $f$ depending on the ULA mass $m$ \cite{Flitter:2022pzf,Ferreira:2020fam}, namely constraints from CMB \cite{Hlozek:2014lca}, BOSS \cite{Lague:2021frh}, SPARC \cite{Bar:2021kti}, Eridanus-II \cite{Marsh:2018zyw}, Lyman-$\alpha$ forest \cite{Kobayashi:2017jcf} and galaxy weak lensing combined with Planck ($+$DES) \cite{Dentler:2021zij}. Constraints from UV luminosity function and optical depth to reionization \cite{Bozek:2014uqa}, as well as constraints from M87 black hole spin \cite{Tamburini:2019vrf, Davoudiasl:2019nlo, Unal:2020jiy}, are implicitly considered, as they do not further reduce the parameter space. Constraints from 21-cm cosmology were not taken into account, as the literature only presents results for $f=1$. In \cite{Flitter:2022pzf} forecasts are obtained by relaxing the assumption on $f$, and it is expected that the Hydrogen Epoch of Reionization Array (HERA) \cite{DeBoer:2016tnn} will be very sensitive to ULDM, comparable to the forecasts for CMB-S4 \cite{Hlozek:2016lzm}. Other constrains on ULAs can arise in specific scenarios, such as \cite{Irsic:2019iff, Fox:2023aat}. Figure \ref{figconst} presents the parameter space and the constraints explicitly considered in this work.

\begin{figure}
\begin{center}
\includegraphics[width=8.5cm]{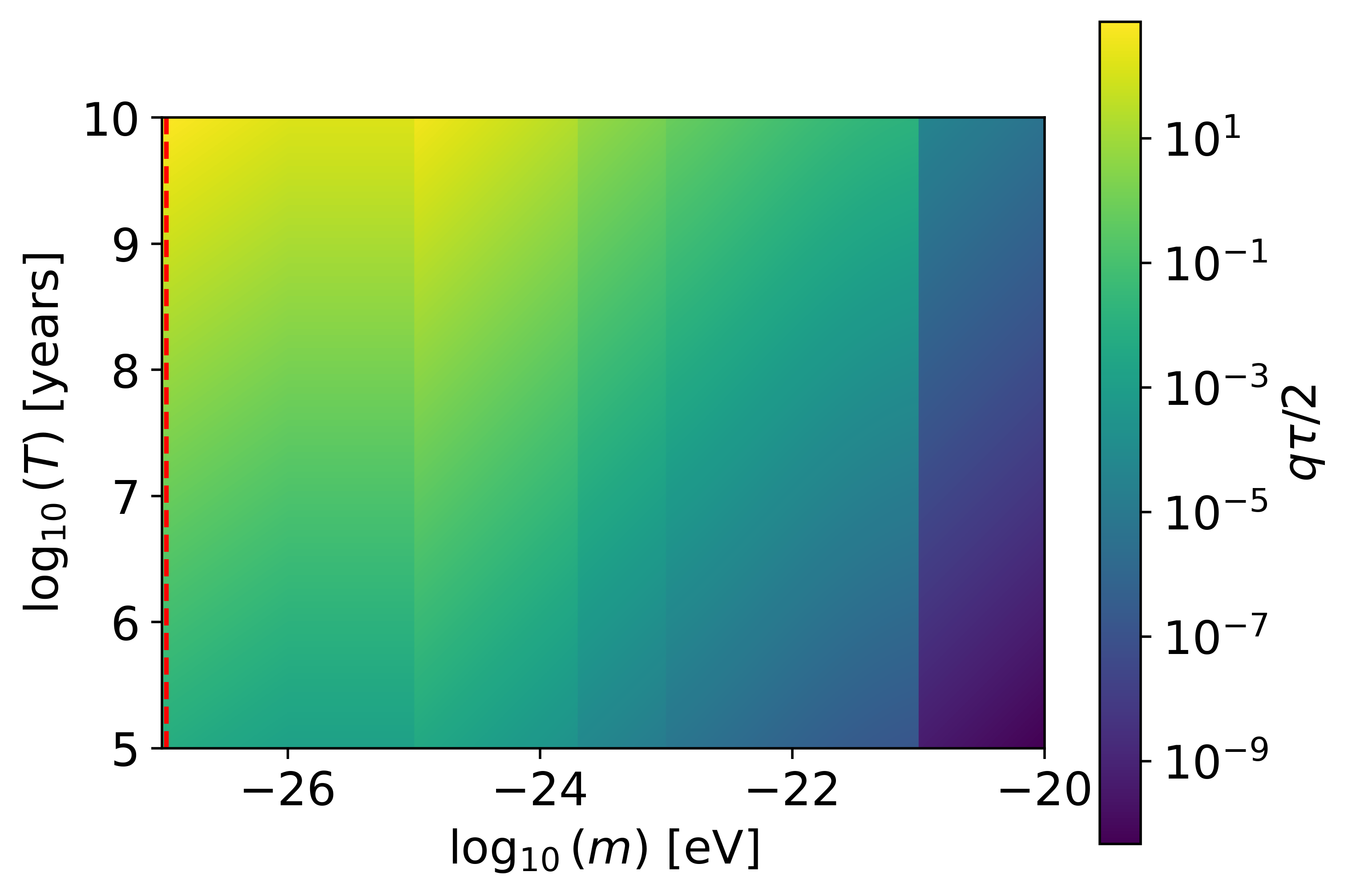}
\end{center}
\caption{\label{figampmassesconst} Argument of the amplification factor \eqref{eqfloquet}, $q\tau/2$, as a function of time and ULA masses, considering $f$ according to the constraints in Figure \ref{figconst}, in the very dense dark matter region, i.e. $1.4\times 10^7$GeV/cm$^3$. The red line indicates when $\rho/m^2\simeq 0.01$, representing a left bound to the region where $\rho/m^2\ll 1$.}
\end{figure}

The Floquet estimates are then computed in the very dense dark matter region, i.e. $\rho_{DM}\simeq 1.4\times 10^7$GeV/cm$^3$, assuming $f$ from the constraints to determine $\rho$ through \eqref{fandrho}. The results obtained for the gravitational wave resonance by considering all the constraints combined are shown in Figure \ref{figampmassesconst}. As expected, the constraints to $f$ suppress $q$ in the Mathieu equation, leading to reduced amplifications, although still significant for some ULA masses. Note that small values of $f$ can be compensated by higher dark matter energy densities $\rho_{DM}$, as long as the latter can be justified in at least one physical regime and $\rho/m^2 \ll 1$. 

\section{Conclusion}\label{secconc}
In this paper I have shown gravitational waves are amplified due to parametric resonance with ULAs constituting the totality or part of a dark matter halo. All the ULA masses considered might only lead to significant amplifications nowadays in very dense dark matter regions, which could exist in different scenarios \cite{Chan:2022gqd, Nampalliwar:2021tyz}. 

Because the gravitational wave resonance could be, in principle, used to independently constrain ULDM (assuming the effect could be measured in a very dense dark matter environment), section \ref{noconst} presents the gravitational wave amplification for $f=1$ for all masses, ignoring existing constraints. The results are depicted in Figure \ref{figampmasses}. On the other hand, in section \ref{secwithconst} the constraints summarized in Figure \ref{figconst} are considered, leading to the gravitational wave amplifications depicted in Figure \ref{figampmassesconst}. Since smaller values of $\rho$ suppress the parameter $q$ in the Mathieu equation \eqref{eqmathieu}, the gravitational wave amplifications are also suppressed compared to $f=1$. A reduced $f$ can be compensated by a denser dark matter region, as long as it can be justified by at least one physical configuration, such as \cite{Chan:2022gqd}, and $\rho/m^2\ll 1$. 

Note that there exists an upper bound on the amplifications established by the condition $h\ll1$, whose value depends on the initial amplitude of the gravitational wave considered. This results from the fact that the Mathieu equation \eqref{eqmathieu} was obtained assuming perturbation theory.  

Given that the parametric resonance occurs for $k=m$, the amplified gravitational waves lie in the range $\sim10^{-8}$Hz to $\sim10^{-13}$Hz, corresponding to $10^{-22}$eV to $10^{-27}$eV. Therefore, the possible gravitational wave sources are primordial perturbations and supermassive black hole binaries. Due to the large time scales for the resonance to become significant, the scenarios of interest should provide gravitational wave emission for a long time, which does happen for the sources mentioned. In addition, note that, although supermassive black hole binaries suffer orbital decay, they present continuous gravitational wave emission long before merger, therefore consistent with the approach carried in this work. Finally, it is relevant to remember that the frequencies in the range $10^{-8}$ Hz to $10^{-9}$ Hz lie in the PTA range, with NANOGrav recently reporting the evidence for a gravitational wave background \cite{NANOGrav:2023gor}.

Throughout the whole paper I have assumed General Relativity and a minimal coupling to the ULAs. Different results could be obtained in modified scenarios such as \cite{Fujita:2020iyx}, where gravitational wave resonance is obtained in the scope of dynamical Chern-Simons gravity. 

Appart from ULAs, other phenomena can trigger the resonance, such as fast small oscillations in the Hubble parameter \cite{Ye:2023xyr} or a varying gravitational wave speed \cite{Cai:2020ovp}.

The consideration of a gravitational wave background might further enhance the resonance \cite{PhysRevD.57.4651, Zanchin:1998fj, Brandenberger:2022xbu}. This investigation is left for future work.\\\\

\begin{acknowledgments}
I would like to thank Shingo Akama, Lucca Fazza, Elisa Ferreira, Alexander Ganz, Chunshan Lin and George Zahariade for useful comments and/or discussions. P.C.M.D. is supported by the grant No. UMO-2018/30/Q/ST9/00795 from the National Science Centre, Poland.\\\\

\end{acknowledgments}

$^\dagger$\href{mailto:paola.moreira.delgado@doctoral.uj.edu.pl}{paola.moreira.delgado@doctoral.uj.edu.pl}\\

\bibliography{main}

\end{document}